\documentclass[8.5pt,twoside,onecolumn]{article}
\usepackage[super,sort&compress,comma]{natbib} 
\usepackage{mhchem}
\usepackage{times}

\usepackage{color}

\usepackage{sectsty}
\usepackage{balance} 
\usepackage{graphicx} 
\usepackage{lastpage}
\usepackage[format=plain,justification=raggedright,singlelinecheck=false,font=small,labelfont=bf,labelsep=space]{caption} 
\usepackage{fancyhdr}

\begin{document}

\thispagestyle{plain}
\fancypagestyle{plain}{
\renewcommand{\headrulewidth}{1pt}}
\renewcommand{\thefootnote}{\fnsymbol{footnote}}
\renewcommand\footnoterule{\vspace*{1pt}%
\hrule width 3.4in height 0.4pt \vspace*{5pt}} 
\setcounter{secnumdepth}{5}

\makeatletter 
\def\subsubsection{\@startsection{subsubsection}{3}{10pt}{-1.25ex plus -1ex minus -.1ex}{0ex plus 0ex}{\normalsize\bf}} 
\def\paragraph{\@startsection{paragraph}{4}{10pt}{-1.25ex plus -1ex minus -.1ex}{0ex plus 0ex}{\normalsize\textit}} 
\renewcommand\@biblabel[1]{#1}            
\renewcommand\@makefntext[1]%
{\noindent\makebox[0pt][r]{\@thefnmark\,}#1}
\makeatother 
\renewcommand{\figurename}{\small{Fig.}~}
\sectionfont{\large}
\subsectionfont{\normalsize} 

\fancyfoot{}
\fancyhead{}
\renewcommand{\headrulewidth}{1pt} 
\renewcommand{\footrulewidth}{1pt}
\setlength{\arrayrulewidth}{1pt}
\setlength{\columnsep}{6.5mm}
\setlength\bibsep{1pt}

\twocolumn[
  \begin{@twocolumnfalse}
\noindent\LARGE{\textbf{Modeling the Behavior of Confined Colloidal Particles Under Shear Flow}}
\vspace{0.6cm}

\noindent\large{\textbf{F.E.~Mackay,\textit{$^{a}$} K. Pastor,\textit{$^{a}$} M. Karttunen, \textit{$^{b}$} and C.~Denniston\textit{$^{a}$} }}\vspace{0.5cm}

\noindent\textit{\small{\textbf{Received Xth XXXXXXXXXX 20XX, Accepted Xth XXXXXXXXX 20XX\newline
First published on the web Xth XXXXXXXXXX 200X}}}

\noindent \textbf{\small{DOI: 10.1039/b000000x}}
\vspace{0.6cm}

\noindent \normalsize{
We investigate the behavior of colloidal suspensions with different volume fractions confined between parallel 
walls under a range of steady shears.  We model the particles using molecular dynamics (MD) 
with full hydrodynamic interactions implemented through the use of a lattice-Boltzmann (LB) fluid.  A quasi-2d ordering occurs in systems characterized by a coexistence of coupled layers with different densities, order, and granular temperature.  We present a phase diagram in terms of shear and volume fraction for each layer, and demonstrate that particle exchange between layers is required for entering the 
disordered phase. 
}
\vspace{0.5cm}
 \end{@twocolumnfalse}
  ]

\section{Introduction}

\footnotetext{\textit{$^{a}$~Department of Applied Mathematics, The University of Western Ontario, London, Ontario N6A 5B8, Canada; E-mail: fmackay2@uwo.ca, cdennist@uwo.ca}}
\footnotetext{\textit{$^{b}$~Department of Chemistry \& Waterloo Institute for Nanotechnology, University of Waterloo, Waterloo, Ontario N2L 3G1, Canada: E-mail: mkarttu@gmail.com}}


Confined particle suspensions are found in a range of 
applications 
including surface coatings and lubricants.  A thorough understanding of 
their rheological 
properties 
is important as the functionality of these products is 
highly dependent on their response to shear.  In a bulk system, the magnitude 
of the applied shear 
dictates the resulting steady state 
behavior.  For 
shear rates above a critical value, any underlying order 
is destroyed, 
and a melted steady state configuration arises.~\cite{ack1981,ste1991,der2009}  In 
contrast, at lower shear rates a reentrant ordering may occur in which particles 
rearrange themselves into hexagonally ordered layers aligned along the 
flow.~\cite{ack1984,dux1996,haw1998,yeo2010,ras1996}  
Such shear alignment 
can be used as a mechanism for manufacturing colloidal 
crystals with minimal imperfections, desirable for photonic band gap 
materials,~\cite{amo2000,kan2005} since shear can both speed up the crystallization process as well as greatly improve the quality of the growing crystal.~\cite{cer2008}  
In the presence of confining geometries, a higher degree of ordering tends to 
form near the walls compared with the sample center,~\cite{haw1998,yeo2010} while packing 
constraints may lead to the formation of new structures not observed in 
bulk.~\cite{coh2004,rei2013}  When shear is applied to colloidal glasses, the resulting behavior provides insight into the flow properties of yield stress materials.~\cite{bes2007, kou2012}

The mechanisms responsible for shear induced structural transitions are not 
fully understood.  Compared with a quiescent sample, both melting and crystallization 
under shear appear to proceed quite differently.~\cite{eis2010}  Based on confocal microscopy, 
Wu et al.~\cite{wu2009} observed that shear induced crystallization 
advances through a collective reordering of the sample, while they suggest that the 
melting process, which is accompanied by large fluctuations in crystalline order, 
involves the nucleation of local domains alternating between disordered and ordered 
states.  Brownian dynamics simulations of a pair of layers driven past one another 
also hint at this latter behavior, with the layers undergoing cycles of order and 
disorder.~\cite{das2002}  However, these simulations ignored 
hydrodynamics and did not allow for  particle exchange between the layers,
which may be 
important for the melting process.  For instance, Palberg et al.~\cite{pal2003} noted 
that the systems they observed first melted in the layer-perpendicular direction and 
with a different mechanism than in-plane melting, although they did not study this 
behavior in detail.  In addition, hydrodynamic interactions can play an important role
in the resulting behavior.  For example, at lower volume fractions than those considered here, 
a balance between hydrodynamic and
interparticle interactions leads to the formation of log-rolling particle strings normal
to the plane of shear.~\cite{che2012,lin2014}

\begin{figure}
\centering
\includegraphics[width=7.4cm]{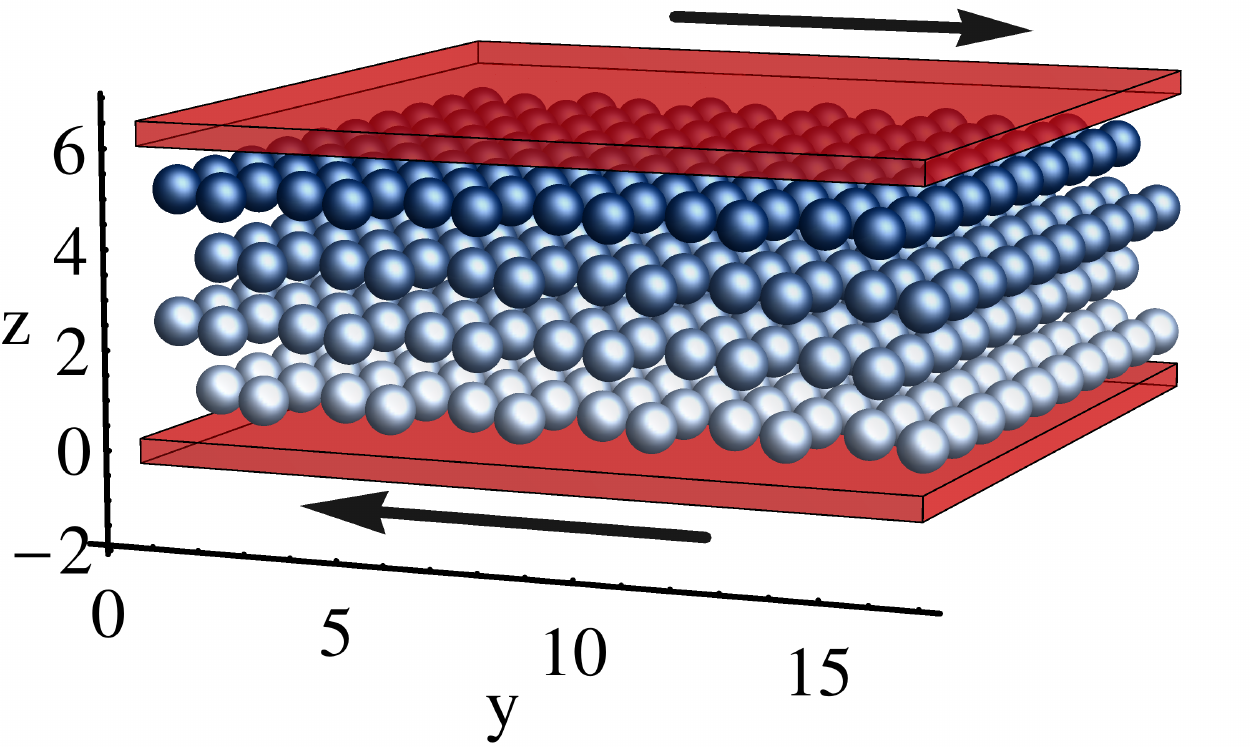}
\caption{Initial system, composed of four layers 
each 
of 120 particles, 
arranged in a hexagonal array. Particle size has been reduced for visualization; 
a range of particle sizes was used.  
Moving walls are present
at $z=0$, and $z=5.88 \mu m$.
}
\label{fig:system}
\end{figure}

To gain further insight into shear induced melting and crystallization, 
we performed hybrid LB-MD simulations of highly confined colloidal systems 
subject to shear.  In contrast to the majority of previous simulations which tend to 
focus exclusively on the colloidal particles, either by not including the effects of 
hydrodynamic interactions~\cite{but1995,mes2006} or employing techniques such as the 
force coupling method,~\cite{yeo2010} which do not directly yield fluid data, 
our simulations 
not only 
track the individual particle motions 
but also 
produce fluid flow data on the entire simulation domain. 

\section{Simulation Details}
Our systems contained 480 colloidal particles confined in a region of 
$16.56 \times 16.56 \times 5.88 ~\mu m$.  Parallel walls were present in the 
$z$-direction, confining the particles into 4 layers.  
Periodic boundary conditions were applied in $x$- and $y$-directions.   
Larger 8 layer systems were simulated producing results similar to the 4 layer case.  
Hence, for  
efficiency, we 
focussed 
on the 4 layer systems.  
Figure~\ref{fig:system} shows the 
initial configuration.
For simplicity, 
we considered a fluid with viscosity and density corresponding to water.  
Shear 
was established 
by moving the upper and lower  
walls at constant, 
opposite velocities, $\pm v_y \hat{{\bf y}}$.  Particle radii and wall 
velocities were varied to investigate a range of shear rates and colloidal volume fractions.

Simulations were performed using LAMMPS~\cite{pli1995} 
along with 
a lattice-Boltzmann fluid package.~\cite{mac2013a}  The latter uses a discretized version of the Boltzmann equation to solve for the fluid motion governed 
by the Navier-Stokes equations,
\begin{eqnarray}
\partial_t \rho + \partial_{\beta} \left( \rho u_{\beta}\right) &=& 0 \nonumber \\
\partial_t\left( \rho u_{\beta} \right) + \partial_{\beta} \left( \rho u_{\alpha}u_{\beta} \right) &=& - \partial_{\alpha}P_0 + F_{\alpha} + \nonumber \\
&& \eta \partial_{\beta}\left( \partial_{\alpha} u_{\beta} + \partial_{\beta}u_{\alpha}\right),
\end{eqnarray}
on a uniform grid.  
Here, $\rho$ is the fluid density, 
${\bf u}$ is the local fluid velocity, $P_0$ is the pressure set to 
$\rho/3 (\Delta x^2/\Delta t^2)$, ${\bf F}$ is a local, external force
resulting from the presence of the colloidal particles, and $\eta$ is the shear viscosity.  
For all simulations, 
we used
$\Delta x = 0.06~\mu m$ for the LB grid spacing, 
and a timestep of $\Delta t = 0.0006~\mu s$.

Each of the colloidal particles was represented by a spherical shell 
of 3612 evenly distributed MD particles, referred to as nodes.  This enables the 
colloidal particles to be modeled off lattice, using molecular dynamics techniques.  
Here, we made use of the rigid fix in LAMMPS, constraining each set of nodes to move 
and rotate together as a single rigid body.  Coupling between the colloidal particles 
and the fluid is then accomplished through the use of conservative forces applied locally to 
both the particle nodes and the fluid according to
\begin{equation}
F = \pm \gamma \left({\bf v} - {\bf u}\right),
\end{equation}
where the $+$ sign corresponds to the force of the node on the fluid, and the $-$ sign 
corresponds to the force of the fluid on the node.  Here, ${\bf v}$ is the velocity of the node, ${\bf u}$ is
 the local fluid velocity, and $\gamma$ is given by
\begin{equation}
\gamma = \frac{2 m_u m_v}{m_u + m_v}\frac{1}{\Delta t_{collision}},
\end{equation} 
where $m_v$ is the mass of the node, $m_u$ is the mass of the fluid, and the collision time, $\Delta t_{collision}$, 
is set equal to the relaxation time in the lattice-Boltzmann fluid.  
  These forces were calculated from the impulses that 
would arise during an elastic collision between the MD nodes and a mass of fluid at 
the node location.  Full details, along with numerous tests of this method can be found in 
Mackay et. al. ~\cite{mac2013a,mac2013b}

In addition to the hydrodynamic forces, we implemented 
interactions between the colloids
by assigning an additional MD particle to each colloidal center interacting via the
truncated \& shifted Lennard-Jones potential 
\begin{equation}
V(r) = \left\{ \begin{array}{ll}
4\epsilon \left[\left(\frac{\sigma}{r}\right)^{12} - \left(\frac{\sigma}{r}\right)^6 \right] - \epsilon,& \text{if } r \leq r_c\\
0,& \text{if } r > r_c,
\end{array}\right.
\end{equation}
where $\sigma = r_c/2^{1/6}$.  
These 
interactions are important
to prevent particle overlap.  We chose $\epsilon = 10$ and set the cutoff 
distance to $r_c = 2(a + \Delta x)$, where $a$ is the MD node placement radius, 
in order to prevent both an overlap of the MD nodes and their associated 
LB grid representations.  A similar interaction was also implemented between 
the central MD nodes and the walls at $z=0$, and $z=5.88\mu m$.  For this 
interaction we set $\epsilon = 20$, and $r_c = a + 3\Delta x$ ensuring 
that no particles left the simulation domain.

Since the resulting colloidal behavior can be quite disordered at times, we make use of the colloidal granular temperature~\cite{granu}, $T$, in order to characterize the colloidal fluctuations.  For a given layer of particles in the system, this temperature is calculated according to
\begin{equation}
T = \frac{1}{3}\left( \left< \left( v_x - \overline{v_x} \right)^2 \right> + \left< \left(v_y - \overline{v_y}\right)^2 \right> + \left< \left( v_z - \overline{v_z} \right)^2 \right>\right),
\end{equation}
where $\overline{v_i}$ is the average velocity of the colloidal particles in the layer, and
the angular brackets correspond to an average over all particles in the layer.  A similar calculation
is also performed for the fluid, proportional to the kinetic energy per unit mass associated with velocity
deviations in the fluid.

\section{Results}
\begin{figure}
\centering
\includegraphics[trim=2.0cm 0cm 3cm 0cm, clip=true, width=\linewidth]{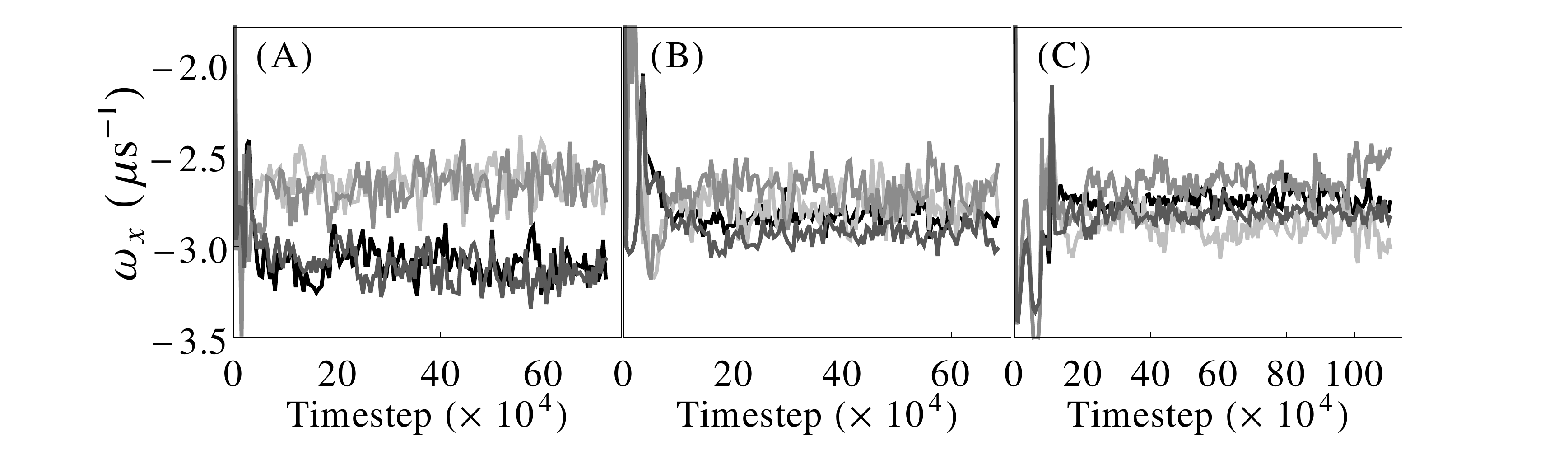}
\caption{
Average angular velocity in each particle layer as a function of time for systems with a shear rate of $6.8 ~\mu s^{-1}$ and volume fractions of (A) 0.521, (B) 0.588 and (C) 0.645. 
Darker grey: particle layers near the walls. 
Lighter grey: middle layers.}
\label{fig:omega}
\end{figure}

Simulations were performed for three different particle radii, corresponding to 
total 
volume fractions of 0.521, 0.588, and 0.645, along with shear rates 
of 1.7, 3.4, 5.1 and 6.8 $\mu s^{-1} \hat{{\bf y}}$.  We chose 
these
high shear rates in order to focus on the effects of hydrodynamic interactions; the shear
in our systems results in colloidal granular temperatures much greater than $k_B T$, isolating
the hydrodynamic effects from those due to thermal fluctuations. 
This shear results in the particles
rotating along the $x$-direction, as illustrated in Figure~\ref{fig:omega}.  However, depending on the shear rate and volume fraction in the system, not all particle layers rotate with the same average rate.  This can be seen in Figure~\ref{fig:omega} (A) where the two (disordered) middle layers exhibit a smaller average angular velocity than the outer layers.

The initial 
configuration chosen does not favour shear along the $y$-direction; 
while shear along $x$ would simply allow the hexagonal layers to slide over one another, 
shear directed along $y$ requires particles in adjacent layers to ``hop'' over one 
another.  Therefore, this initial particle arrangement is not stable against shear 
and the systems evolve into a variety of different configurations depending on the 
volume fraction and shear rate. 

For the majority of systems considered, we observed a migration of particles from the 
middle layers towards the walls, consistent with previous works.~\cite{yeo2010} 
This leads to a range of different layer volume fractions.  
We therefore focus 
on the per layer behavior of our systems by dividing the system along the vertical direction into distinct regions of equal size, and assigning particles to each of these layers based on the location of their center of mass.  We then quantify the behavior in each layer through the local 2D 
orientational order parameter
\begin{equation}
\psi_6(r_{jk}) = \frac{1}{n} \sum_{k=1}^n e^{i6\theta(r_{jk})},
\end{equation}
averaged over all the particles in each layer.
Here, $n$ corresponds to the number of nearest neighbours 
of
particle $j$, calculated as the particles, $k$, in the same layer as $j$, with a surface to surface separation from $j$ within a particle diameter, and $\theta(r_{jk})$ is the angle between the vector connecting 
particle $j$ with its neighbour $k$ and an arbitrary fixed axis.  
For a perfectly ordered hexagonal system, $\left<|\psi_6| \right> = 1$.  

\begin{figure}
\centering
\includegraphics[width=7cm]{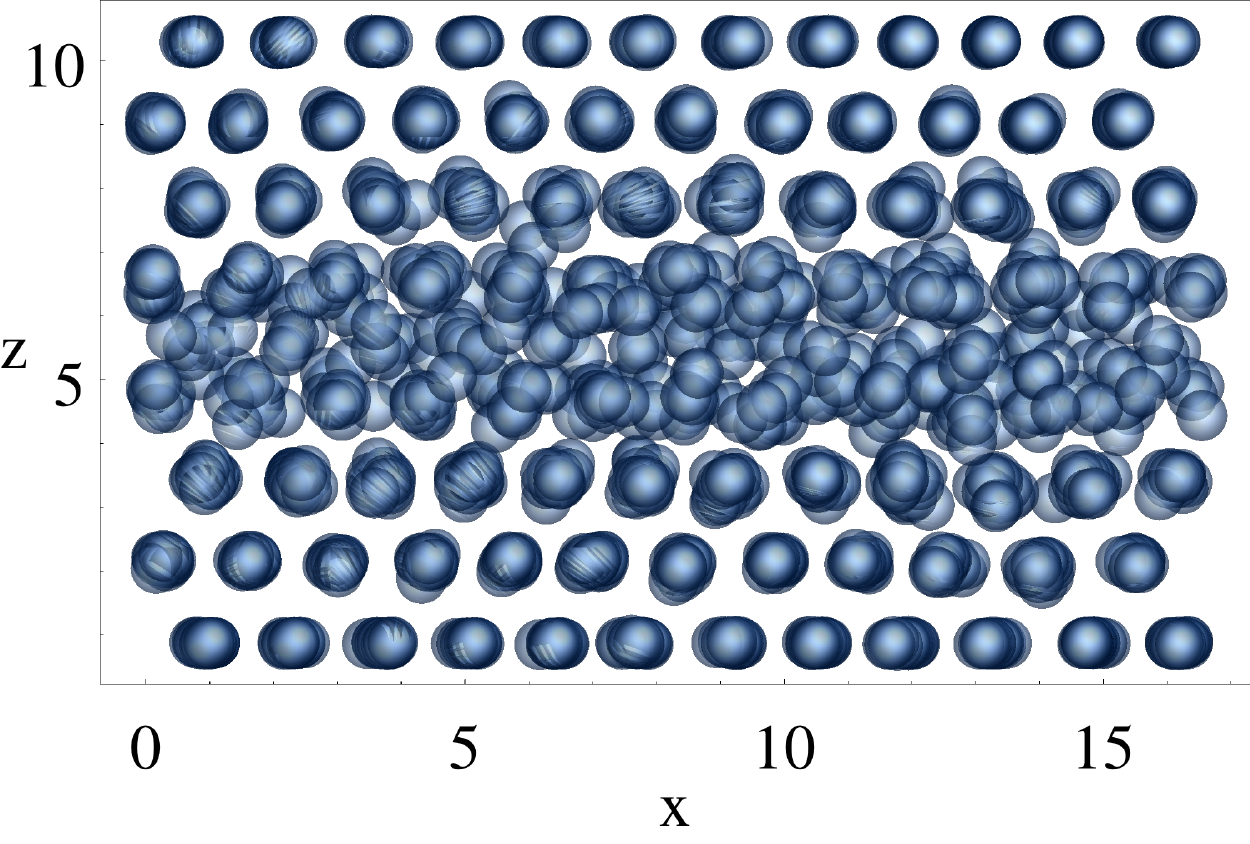}
\caption{
Snapshot in the $x$-$z$ plane of an initially ordered eight layer system which has phase separated into six ordered outer layers, and two disordered middle layers.  Note that the size of the colloidal particles has been reduced here for clarity.
}
\label{fig:eight}
\end{figure}

Simulations were run until steady state was achieved and the order parameter amplitude
in each layer was observed to fluctuate around a constant value.  
The layers of the resulting systems are 
characterized as disordered when $\left<|\psi_6| \right> < 0.6$, or hexagonally ordered 
with alignment now along the flow direction when $\left<|\psi_6| \right> > 0.6$. 
For high volume fractions with low to moderate shear rates, all layers
reorder to form hexagonal particle arrangements aligned along the flow.  
In contrast, at low volume fractions and high shear, the steady state configuration 
consists of four disordered layers.  For the remaining systems, a phase 
separation occurs, leading to the appearance of distinct ordered and disordered layers.  
The formation of distinct layers
as opposed to regions of order and disorder
within a given layer could be 
a finite size effect.
However, this behavior is consistent
with experiments in which hexagonal layers are observed coexisting with a fluid 
phase.~\cite{pal2003,bie2004,wu2009}  Similar behavior was also observed in an 
eight layer system (size $16.56 \times 33.12 \times 11.16 \mu m$), where the 
system phase separated into six ordered and two disordered layers as shown in Figure~\ref{fig:eight}.

\begin{figure}[t]
\centering
\includegraphics[width=8cm]{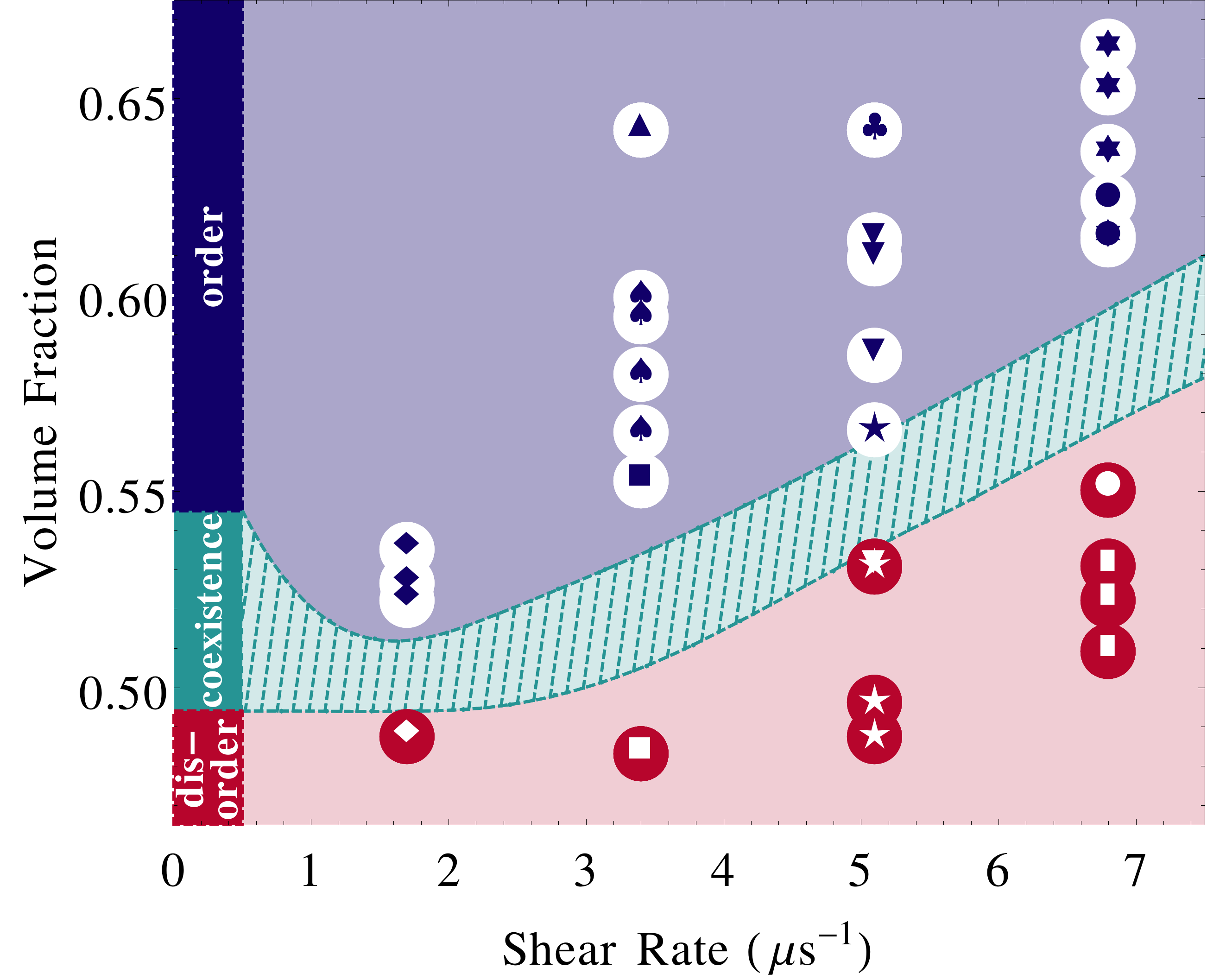}
\caption{Phase diagram illustrating the per layer system behavior as 
a function of volume fraction and shear rate in each layer.  
Markers corresponding to a blue symbol on a white circular background are associated with hexagonally ordered layers, while those with a white symbol on a red background denote disordered layers. At a given shear rate, markers with the same central symbol correspond to layers from the same physical system.  For example, at a shear rate of $1.7 \mu s^{-1}$ there are four markers each with a central diamond symbol.  These correspond to layers from a system with total volume fraction 0.521 which has phase separated into 1 disordered and 3 ordered layers.  Similarly, squares, 5 point stars and rectangles are also associated with systems with a total volume fractions of 0.521, clubs, downward pointing triangles, and circles correspond to a total volume fraction of 0.588, and upward pointing triangles, clovers, and 6 point stars correspond to a total volume fraction of 0.645.
For purposes of comparison the location of the ordered, disordered, and coexistence region for a colloidal system in the absence of shear is indicated on the left.
}
\label{fig:phase}
\end{figure}

\begin{figure*}
\centering
\includegraphics[width=\linewidth]{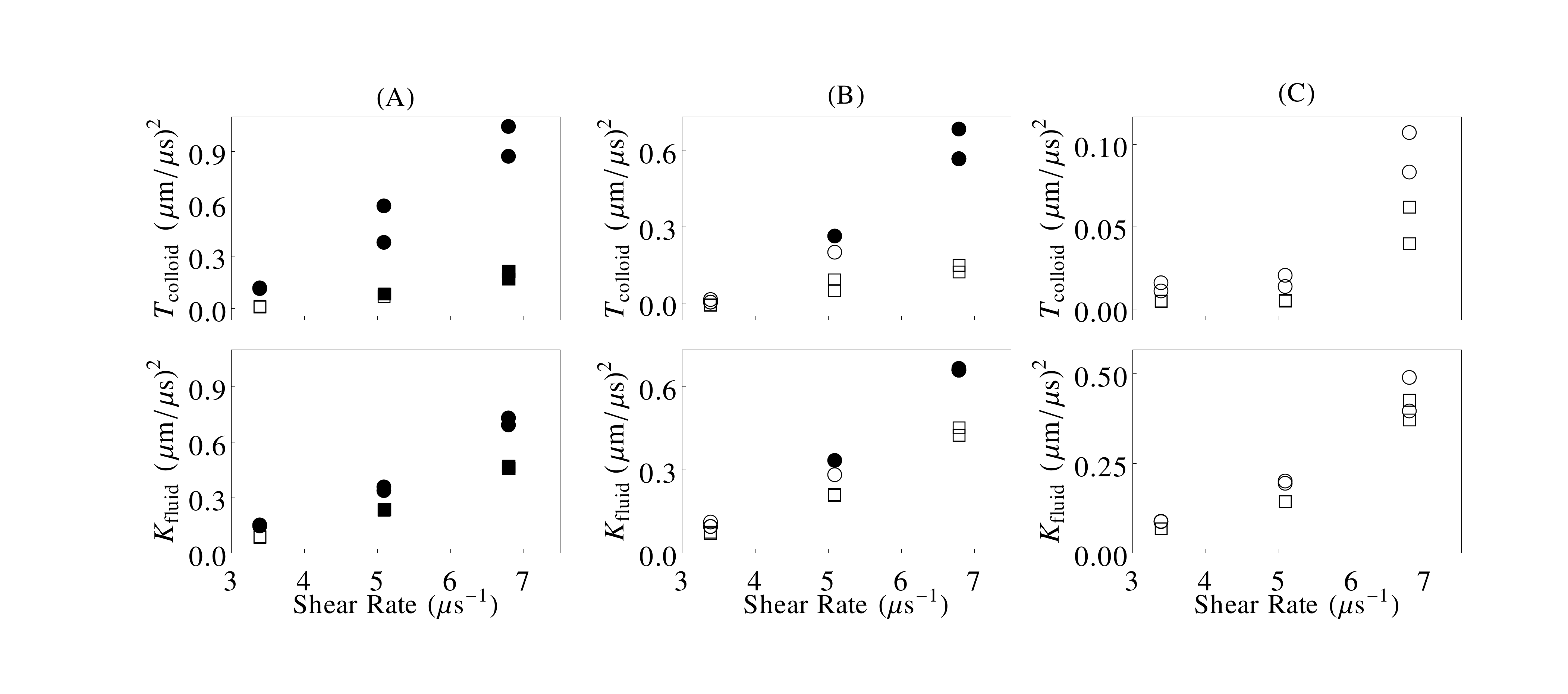}
\caption{Per layer, steady-state granular temperature for the colloids 
(top panel) and kinetic energy per unit mass for the fluid (lower panel).  
(A) Volume fraction of 0.521, (B) 0.588, and (C) 0.645.  
Squares: wall layers. Circles: middle layers. Open symbols 
correspond to ordered and closed to disordered layers.  Note the use of different vertical axes for each plot.
}
\label{fig:temp}
\end{figure*}

Figure~\ref{fig:phase} presents our results in the form of a per layer phase diagram, 
illustrating the layer order as a function of volume fraction and applied shear rate.
As expected, \textit{high shear rates} tend to induce disorder, raising the colloidal melting 
volume fraction above that of a system in the absence of shear. 
Shear alignment acts at \textit{lower shear rates}, enabling ordered layers 
to form at volume fractions reduced from the stationary system value. 
(At even lower shear rates, we expect the melting \& freezing volume fractions 
to return to their zero shear levels; due to the significantly longer simulation 
times 
we have omitted these from our analysis.)  
As Fig.~\ref{fig:phase} shows, the location of the order-disorder phase boundary is mapped out by the systems 
which experience {\it phase separation}.  In these systems, higher volume fraction ordered outer 
layers coexist with lower volume fraction disordered middle layers, where the number of each 
type that develop depends on both 
the location of the phase boundary and the size of the associated coexistence region, 
inaccessible to a pure (ordered or disordered) phase.  For example, consider the system 
shown in the lower left hand corner of Fig.~\ref{fig:phase} consisting of three ordered layers 
and one disordered middle layer.  The asymmetry 
of the system clearly points to 
an inaccessible region on the phase diagram separating the ordered and disordered phases, 
explaining why the two middle layers do not form the same phase with volume fractions 
which would then be located inside this region.  
The ability of these systems to phase separate is enabled only through a migration of 
particles towards the walls, coupled with lower granular temperature, $T$, near the walls 
(see Fig.~\ref{fig:temp}).  These lower observed temperatures arise from a reduction in the degrees of freedom imposed 
by confinement.  Together with the particle migration towards the walls, this leads to the formation of ordered wall layers, 
leaving behind lower volume fraction, disordered middle layers.
Particle migration appears to 
be a general characteristic of the approach towards the phase boundary; moving towards this 
boundary, either by adjusting the volume fraction or the shear rate, results in the 
appearance of a range of layer volume fractions preceding any phase separation.  In contrast, Cohen et. al.~\cite{coh2004} found that  
systems in contact with a reservoir do not exhibit such behavior.  Instead, the balance 
between viscous stress and osmotic pressure sets a high volume fraction throughout the system 
leading to the formation of dense, ordered particle structures.

\begin{figure*}
\centering
\includegraphics[width=\linewidth]{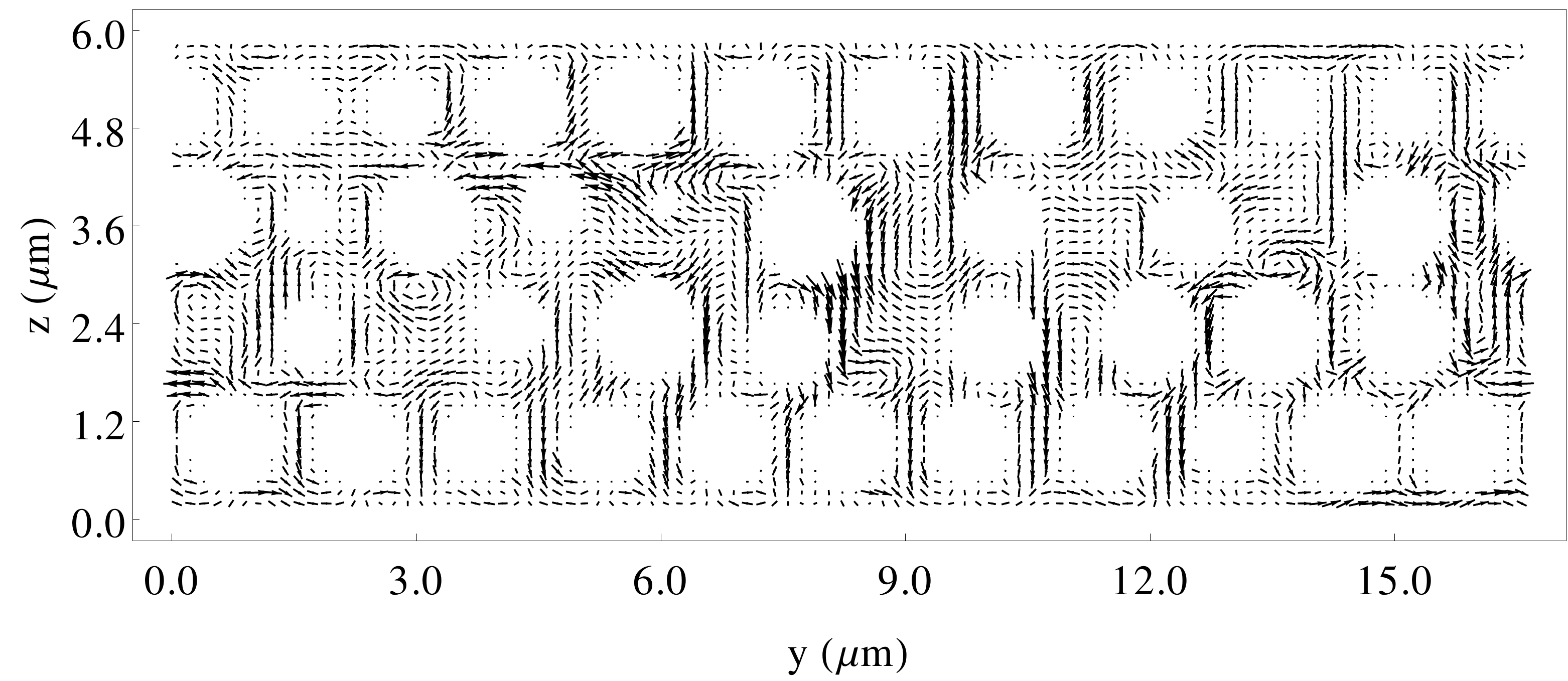}
\caption{Fluid velocity deviations in the $yz$-plane, calculated by subtracting off 
the average fluid velocity at each $z$ position, for a system with volume fraction = 0.588, and shear rate = 6.8 $\mu s^{-1}$. 
Here, the colloidal particles are moving and rotating with the fluid, and the fluid inside them has been removed for
 illustrative purposes.}
\label{fig:velocity}
\end{figure*}

The onset of particle migration is accompanied by a noticeable rise in the colloidal granular temperature.  
Consider, for example, the highest volume fraction systems. These all consist of four 
ordered layers in steady state for the shear rates we have investigated.  Moving from lower 
to higher shear rate systems, we observe a fairly steady rise in the kinetic energy per unit mass for the fluid.
However, the colloidal granular temperatures remain approximately 
constant among the lower shear rate systems, only increasing at the onset of particle 
exchange among layers (see Fig.~\ref{fig:temp}C).  Once the phase boundary is crossed, 
and disordered layers begin to form, we observe an increase in both the colloidal granular temperature and the kinetic energy in the fluid, 
particularly in the disordered layers 
(see Figs.~\ref{fig:temp}A and B).  The velocity deviations in the fluid are illustrated in 
Fig.~\ref{fig:velocity} for one such system.  Here we have subtracted off the average velocity profile in the system.  Clearly, the velocity field can be quite complicated, particularly in the disordered 
middle layers, where the chaotic motion of the colloidal particles can lead to disordered fluid flow around them. 

\begin{figure*}
\centering
\includegraphics[width=\linewidth]{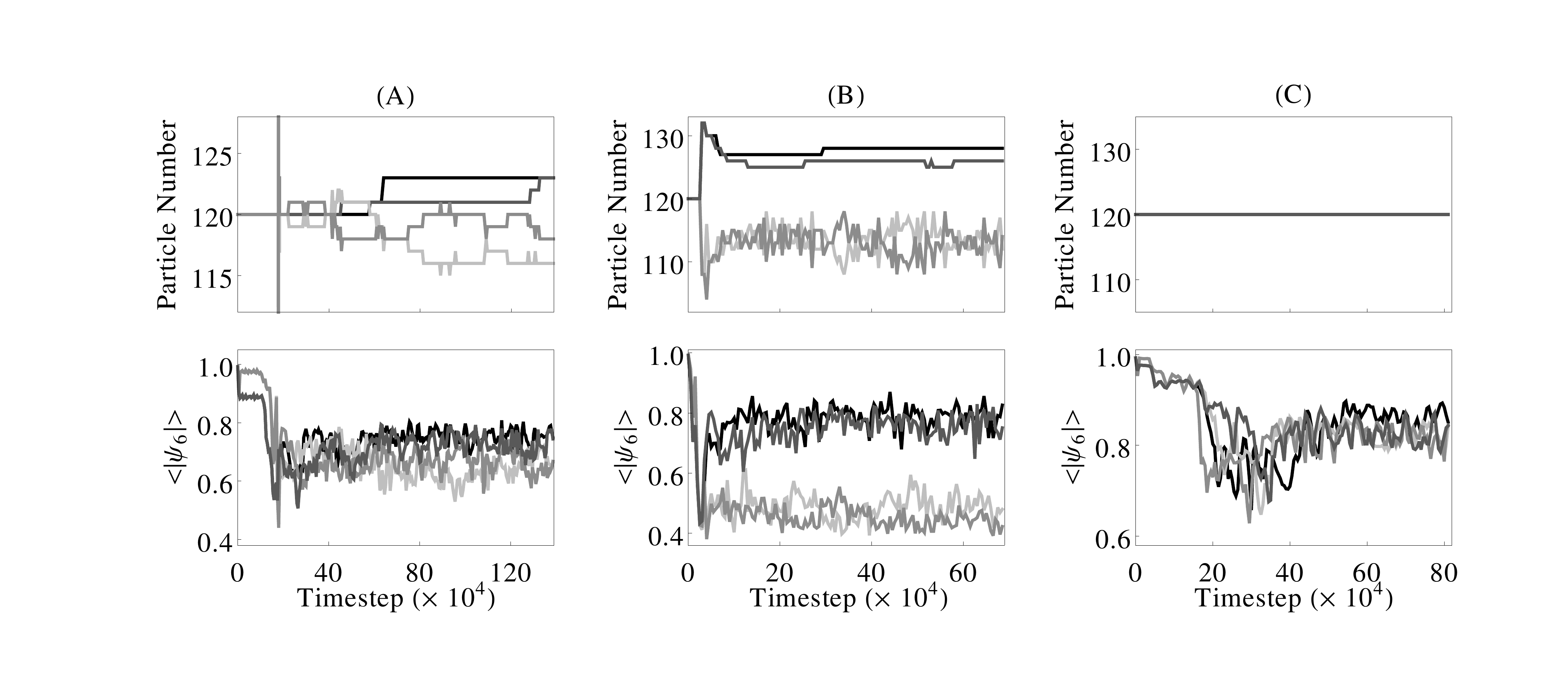}
\caption{Particle number and order parameter amplitude, $\left<|\psi_6| \right>$, 
as a function of time for each layer. 
(A) volume fraction = 0.588, shear rate = 3.4 $\mu s^{-1}$, 
(B) volume fraction = 0.588, shear rate = 6.8 $\mu s^{-1}$,
(C) volume fraction = 0.645, shear rate = 3.4 $\mu s^{-1}$.
Darker grey: particle layers near the walls. 
Lighter grey: middle layers.
}  
\label{fig:denorder}
\end{figure*}

\begin{figure*}
\centering
\includegraphics[width=0.8\linewidth]{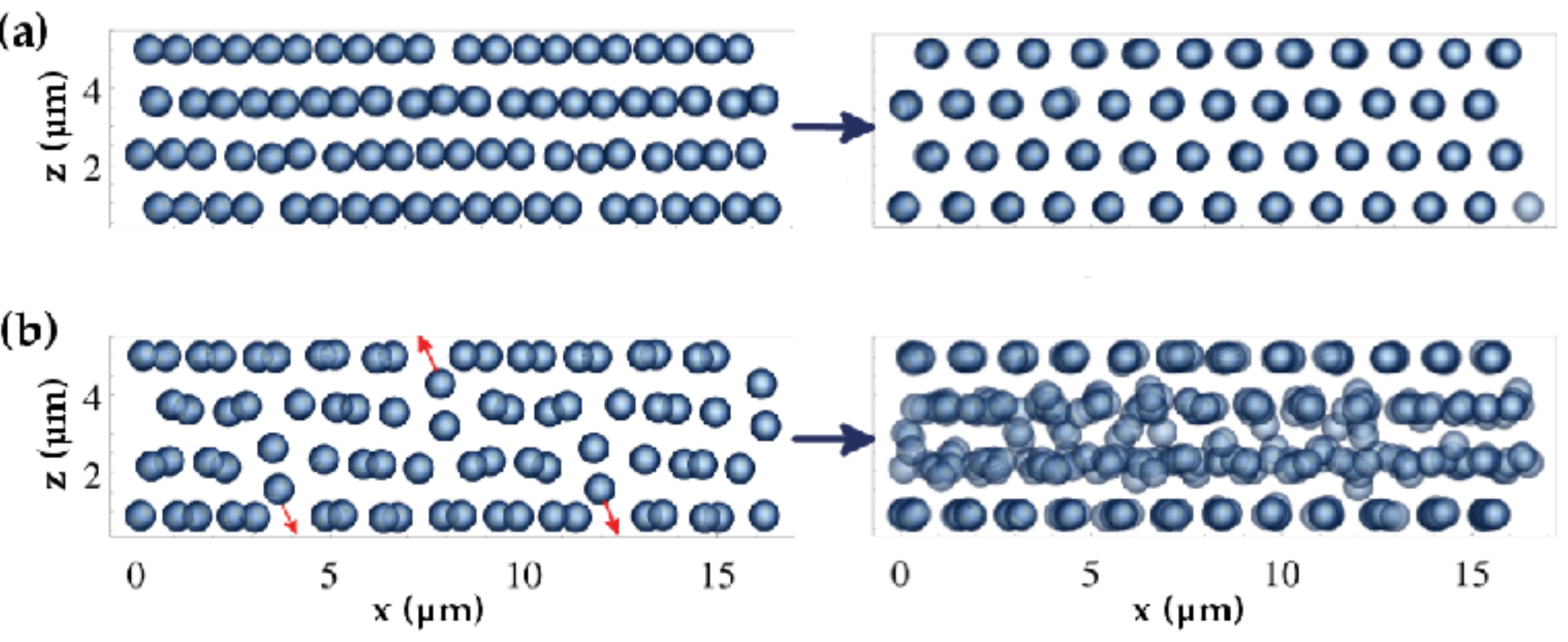}
\caption{Snapshots of configurations in the $x$-$z$ plane, for 
(a) volume fraction = 0.645, shear rate = 3.4 $\mu s^{-1}$, and 
(b) volume fraction = 0.588, shear rate = 6.8 $\mu s^{-1}$.  
Left: early times in the simulation. 
Right: the steady-state configurations.  
Note that the size of the colloidal particles has been reduced here for clarity.
}
\label{fig:melting}
\end{figure*}

To gain a better understanding of the mechanisms driving the order-disorder (and vice-versa) transitions in the layers, we monitor the evolution of the systems as a function of time.
Figure~\ref{fig:denorder} shows the results for three different simulations, 
which we will refer to as systems A, B and C.  Comparing our order parameter to the experimental 
results of Wu et al. ~\cite{wu2009}, we find a qualitatively similar behavior.  
We observe the same initial rapid decrease in order followed by fluctuating values.  For 
system B, in which the two inner layers are disordered, the variations in order are somewhat 
large, ranging from $\sim 0.4 - 0.6$.  As discussed by Wu et al., these fluctuations 
suggest the formation of local ordered domains which regularly appear and subsequently 
re-melt.  This behavior is confirmed in our simulation.  However, the magnitude of our 
fluctuations are smaller than those observed by Wu et al. and are further reduced in our 
lower volume fraction systems.  These fluctuations in layer order are clearly related to 
the fluctuations exhibited by the layer particle number, with increased particle numbers 
resulting in higher levels of order.  In addition, particle exchange between layers appears 
to directly affect the level of order in the systems.  The initial 
drop in order ($\left<|\psi_6| \right>$ decreasing to values below 0.6), as exhibited 
by systems A and B directly corresponds to the emergence of particle exchange. 
For system C, in which no such exchange occurs, $\left<|\psi_6| \right>$ never falls 
below 0.65.  The initial particle exchange appears to proceed as follows: As previously 
mentioned, shear along the $y$-direction is unfavourable to the initial particle configuration, 
requiring particles in adjacent layers to ``hop'' over one another.  This leads to 
particle motion in the $z$-direction, which causes particles to push on adjacent layers 
leading to the formation of voids.  We find this behavior to be exhibited by all of the 
systems, even those which do not experience particle exchange (see, e.g., the 
left panel of Fig.~\ref{fig:melting}a, corresponding to the early time behavior of 
system C).  However, for systems in which the particle size is small enough, and 
the $z$-motion large enough, particles from neighbouring layers are able to move 
into the voids and exchange occurs.  This is illustrated in the left hand panel 
of Fig.~\ref{fig:melting}b, showing the early time behavior of system B.

While systems A and C both reach steady-state configurations consisting of four ordered 
layers, the fact that particle exchange occurs for one system and not the other clearly 
points to different mechanisms responsible for the layer reordering.  Particle exchange 
tends to induce disorder, effectively melting the layers, which subsequently reorder.  
In contrast, in the absence of particle exchange, regions within the layer undergo a 
reordering.  In both cases, the result is the formation of hexagonal particle layers 
aligned along the flow (right hand panel of Fig.~\ref{fig:melting}a).

When particle exchange does occur among layers, aside from the initial drop in order 
that accompanies the onset of exchange, the degree of layer order throughout the 
simulation is directly related to the quantity and frequency of exchange.  For example, 
even though all four layers in system A reorder, a moderate amount of particle exchange 
occurs for the middle layers.  Correspondingly, $\left<|\psi_6| \right>$ is reduced to 
the range $0.6 \leq \left<|\psi_6| \right> \leq 0.7$ compared with the outer layers, 
which experience almost no exchange and have higher values for $\left<|\psi_6| \right>$.  
Even more frequent particle exchange is associated with even lower values for the 
$\left<|\psi_6| \right>$.  For system B, in which the inner layers experience a 
substantial and sustained particle exchange, the order parameter amplitude falls to 
$0.4 \leq \left<|\psi_6| \right> \leq 0.55$, corresponding to disordered layers.  This 
particle exchange, illustrated in the right hand panel of Figure \ref{fig:melting}b, 
persists along with the disorder in the simulations, and is associated with both 
increased colloidal granular temperature and kinetic energy in the fluid.

\section{Conclusions}
In this work, we investigated the effects of volume fraction and shear rate on the behavior of confined colloidal 
particles.  Starting from an initially ordered arrangement, the resulting steady-state configurations consist 
of particles which either reorder into hexagonally ordered layers aligned along the flow, form purely disordered 
layers, or phase separate into higher volume fraction ordered layers near the walls, and lower volume fraction 
disordered middle layers.  By plotting the per layer behavior as a phase diagram, we illustrate the effects 
of the volume fraction and shear stress in our systems on the colloidal order-disorder transition.  
An examination of the layer particle number as a function of time reveals that the onset of disorder in 
the systems is characterized by the emergence of particle exchange among layers: Systems which do not 
experience exchange never pass through the disordered phase.  When particle exchange occurs, 
there is a direct relation between the amount of exchange, and a reduction in the layer order, with 
disordered layers experiencing substantial and sustained particle exchange.  It is important to note that all of these effects are driven by the hydrodynamic flow rather than thermal noise.  At times, the resulting particle motion can be random, disordered and lacking any periodicity, which naturally leads to disordered fluid flow around the particles despite the overall average imposed shear gradient.

\section*{Acknowledgments}
We thank the Natural Science and Engineering Research Council of Canada (NSERC) for financial support.  
This research has been enabled by the use of computing resources provided by WestGrid, SharcNet and Compute/Calcul Canada.

\end{document}